\documentclass[letterpaper,twocolumn,10pt]{article}
\usepackage{usenix2019_v3}
\usepackage[frozencache,cachedir=.]{minted}

\usepackage{graphicx}
\usepackage{subcaption}

\begin{document}


\title{\Large \bf ProTuner: Tuning Programs with Monte Carlo Tree Search}
\author{
  Ameer Haj-Ali\\
 UC Berkeley
 \and
  Hasan Genc\\
 UC Berkeley
 \and
   Qijing Huang\\
 UC Berkeley
 \and
   William Moses\\
 MIT
 \and
   John Wawrzynek\\
 UC Berkeley
 \and
   Krste Asanović\\
 UC Berkeley
 \and
  Ion Stoica\\
 UC Berkeley} 

    \maketitle

\begin{abstract}
We explore applying the Monte Carlo Tree Search (MCTS) algorithm in a notoriously difficult task: tuning programs for high-performance deep learning and image processing. We build our framework on top of Halide and show that MCTS can outperform the state-of-the-art beam-search algorithm. Unlike beam search, which is guided by greedy intermediate performance comparisons between partial and less meaningful schedules, MCTS compares complete schedules and looks ahead before making any intermediate scheduling decision. We further explore modifications to the standard MCTS algorithm as well as combining real execution time measurements with the cost model. Our results show that MCTS can outperform beam search on a suite of 16 real benchmarks. 
\end{abstract}
\section{Introduction}
Most deep learning and image processing programs rely heavily on loops, which represent the vast majority of a program's total execution time. Due to the high number of loops in each loop nest, it is possible to schedule the loops in many different, yet functionally equivalent ways. Choosing a bad program schedule can result in a dramatically worse execution time. Similar schedules can include different optimizations such as inlining, tiling, vectorization, and multithreading, which can significantly impact the performance of the program. The number of possible optimizations grows exponentially with the number of loops in the loop nest. With the end of Moore's law and the booming of new application-specific integrated circuits (ASICs), the scheduling challenge becomes harder as the scheduler also needs to generate different schedules for different target architectures.

Due to its complexity, scheduling is often done using heuristics and tremendous amounts of hand engineering. Big vendors often hire engineers that are dedicated to manually writing schedules for different applications. This incurs huge costs both in terms of time and human capital. Heuristics mostly fall short in achieving optimal performance~\cite{cummins2017synthesizing,haj2020autophase,haj2020neurovectorizer}. Ideally, the scheduler should consider all the possible schedules and the target hardware to find the optimal schedule. Unfortunately,  the space of possible schedules for different hardware targets is prohibitively large to explore. To cope with that, a recent work~\cite{adams2019learning} proposed using beam search with a learned cost model to find good schedules. While this approach achieves promising improvements over the baseline default auto-scheduler in Halide~\cite{ragan2013halide}, it often fails to find the optimal schedule. The main issue lies in the greediness of beam search and the inability of the cost model to accurately predict the performance of partially scheduled (not meaningful) programs, which is needed at every intermediate step of the beam search. This results in a multiplied error at every decision made in the beam search tree and an inability to explore schedules that are less rewarding in the short term but potentially more rewarding in the long term.

To overcome these challenges we propose ProTuner, which uses Monte Carlo Tree Search (MCTS)~\cite{browne2012survey}. In ProTuner, we formulate the scheduling problem as a Markov decision process (MDP) where each intermediate schedule is represented as a state and the actions are the different intermediate optimizations that could be applied next. The reward is proportional to the execution time improvement. The solution would be the actions that lead to the optimal schedule and hence solving the MDP can guarantee the optimal schedule. One promising algorithm to solve MDPs is MCTS. MCTS builds a search tree using selection, expansion, simulation, and backpropagation that explore the search space. After some number of iterations, the tree decides which next step (of intermediate scheduling optimization) to perform next. When this happens, a new root is determined and the MCTS starts again. With the upper confidence bound (UCB)~\cite{auer2002finite} the MCTS is guaranteed to converge to the optimal solution after enough iterations.

MCTS makes decisions by looking ahead, evaluating complete schedules, avoiding greediness, and considering the expected long term reward of scheduling decisions, which also makes it more resilient to noise in the cost model. We, therefore, conjecture that MCTS is a better fit for finding the optimal schedule compared to beam search. To explore that, we implemented ProTuner on top of Halide~\cite{ragan2013halide}, evaluated it, and found that ProTuner with MCTS outperforms beam search or achieves comparable performance on all of the evaluated benchmarks, achieving up to $3.25\times$ better performance. 

Our main contributions are:
\begin{itemize}
    \item We propose to formulate the scheduling problem as an MDP and solve it using MCTS with the UCB.
    \item We build ProTuner on top of Halide and explore different MCTS techniques to improve and fine tune its performance.
    \item Rigorous evaluations that show ProTuner achieves up to $3.5\times$ better performance than beam search on a suite of 16 real benchmarks.
    \item We show how ProTuner can combine real execution time evaluation with the learned cost model and show this can further boost the performance.
\end{itemize}
\section{Background}
\subsection{Halide Scheduling}
Halide~\cite{ragan2013halide} is a domain-specific language for image processing and deep learning tasks. 
Halide's language abstraction decouples the algorithmic descriptions of the target image processing workloads from a specific mapping of the workload on hardware, which we refer to as a ``schedule''. This abstraction provides the user with a clearly defined scheduling space and makes it easier to explore different schedules automatically.
Many decisions need to be made in a Halide schedule, including the execution order of different functions, vectorization factors, tiling factors, inlining, memory allocation strategies, etc.
The overall scheduling space is intractable and expert scheduling can be hard to develop. Therefore, automatic generation of high-performance Halide schedulings has been implemented and studied in several prior works  ~\cite{mullapudi2015polymage}\cite{mullapudi2016automatically}\cite{sioutas2019schedule}\cite{adams2019learning}.

\subsection{Beam Search}
Beam search~\cite{reddy1977speech} is a heuristic algorithm that explores a decision tree and searches for the optimal decisions by expanding a limited number of children with the highest intermediate rewards. It is widely used for the sequential decision-making process, such as speech recognition~\cite{reddy1977speech} and software scheduling~\cite{adams2019learning}. It builds its search tree with a breadth-first search. 
At each step of the algorithm, it exhaustively evaluates all the direct children, sorts the children based on the intermediate rewards, and keeps the top-$k$ children as the parent nodes for the next iteration. 
$k$ is the beam size that determines the total number of top children to keep at every iteration.
It is essentially a greedy algorithm and thus can get stuck in local optima. 

\subsection{Markov Decision Processes}
A Markov decision process (MDP) is a discrete stochastic control process that models sequential decision making in fully observable environments. It assumes the Markov property that the impact of one decision taken in a state only depends on that state and not the prior decision history. 
An MDP model consists of:
\begin{itemize}
    \item $S$: A set of possible states, with $s_0$ representing the initial state. 
    \item $A$: A set of possible actions. 
    \item $R(s, a)$: A reward function. 
    \item $T(s' | s, a)$: A transition function that models the probability of getting to state $s'$ given an action $a$ in state $s$. 
\end{itemize}
Solving an MDP means finding a policy $\pi(s)$ that chooses an action to apply based on the current state and optimizes for the overall expected reward. The decision making process is modeled as a sequence of state and action pairs $(s, a)$. The next state $s'$ can either be decided deterministically by the pair $(s, a)$ or stochastically by a probability distribution $p(s'|s, a)$.  


\subsection{Monte Carlo Tree Search}
Monte Carlo Tree Search (MCTS) is a method that can solve MDPs. It combines tree search with random sampling for finding the optimal decisions in the MDP. In MCTS, a tree is built incrementally based on selection, expansion, simulation, and backpropagation. 
A \textit{tree policy} is used to \textit{select} a node to expand at each iteration of the algorithm. This policy should balance the exploration and exploitation of the search algorithm. The node is \textit{expanded} and a \textit{simulation} is then run from the selected node to collect the rewards of the terminal state. The decisions made during the simulation are determined by a \textit{default policy}, which can be uniform random sampling in its simplest form and this is what we use. Lastly MCTS backpropagates the reward and updates the statistics of the ancestor nodes. In this paper, the tree policy used is the UCB~\cite{auer2002finite}:
\begin{equation}
    UCB = \bar{X_j} + 2C_p\sqrt{\frac{2ln(n)}{n_j}}
\end{equation}
where $n$ is the number of times the current parent node has been visited, $n_j$ is the number of times child $j$ has been visited, and $C_p>0$ is a constant. $\bar{X_j}$ is the average reward of the simulations. The left term ($\bar{X_j}$) tracks exploitation while the right term tracks exploration. Increasing $C_p$ will add more exploration, and decreasing it will reduce exploration.

According to \cite{jakl2011arimaa}, MCTS offers significant advantages over alpha-beta pruning that minimizes the search space in the scenario where there is no good evaluation function. The evaluation of moves in MCTS has been proven to converge to Minimax.


\begin{figure*}[!t]
    \centering
    \includegraphics[trim={0cm 1cm 0cm 0cm},width=\textwidth,clip]{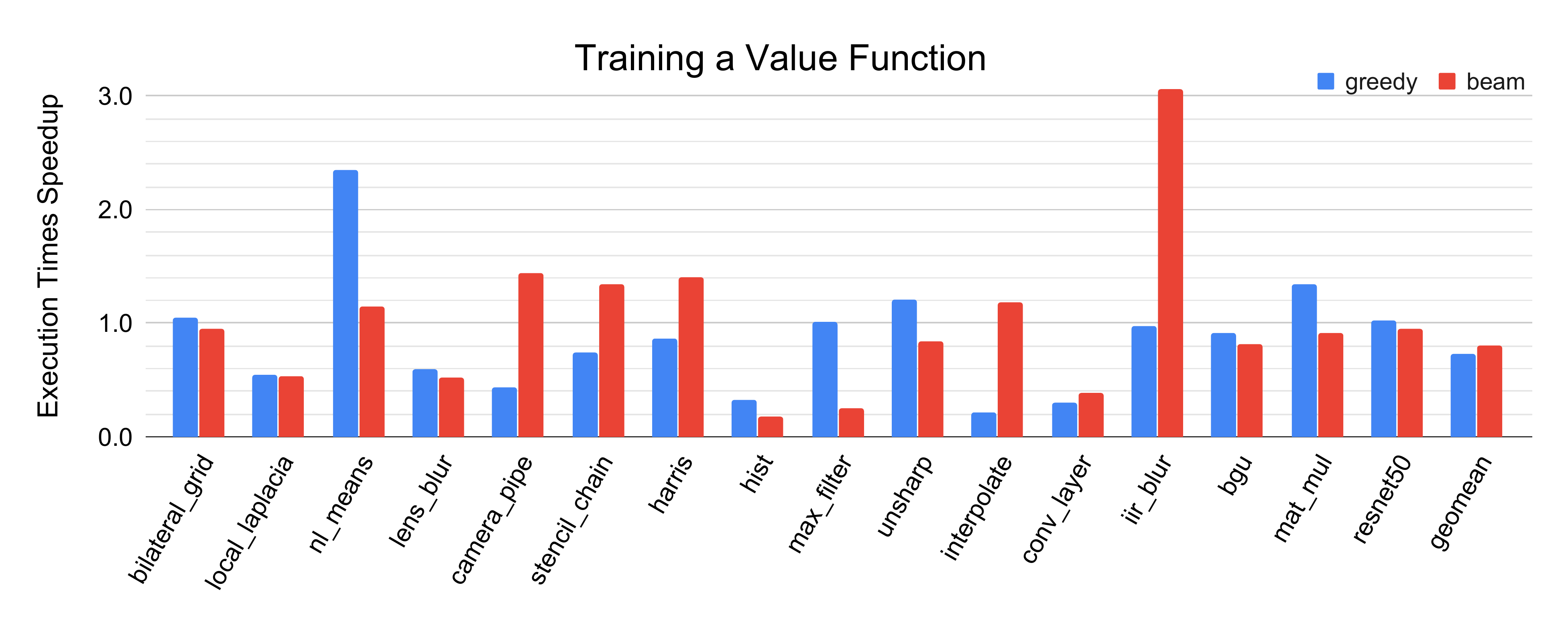}
    \caption{Speedup of greedy and beam search with a cost model trained to predict the cost of the future complete schedules normalized to greedy and beam search respectively with the original cost model (trained on complete schedules only).}
    \label{fig:valuefunc}
\end{figure*}

\begin{figure*}[!t]
    \centering
    \includegraphics[trim={0cm 1cm 0cm 0cm},width=\textwidth,clip]{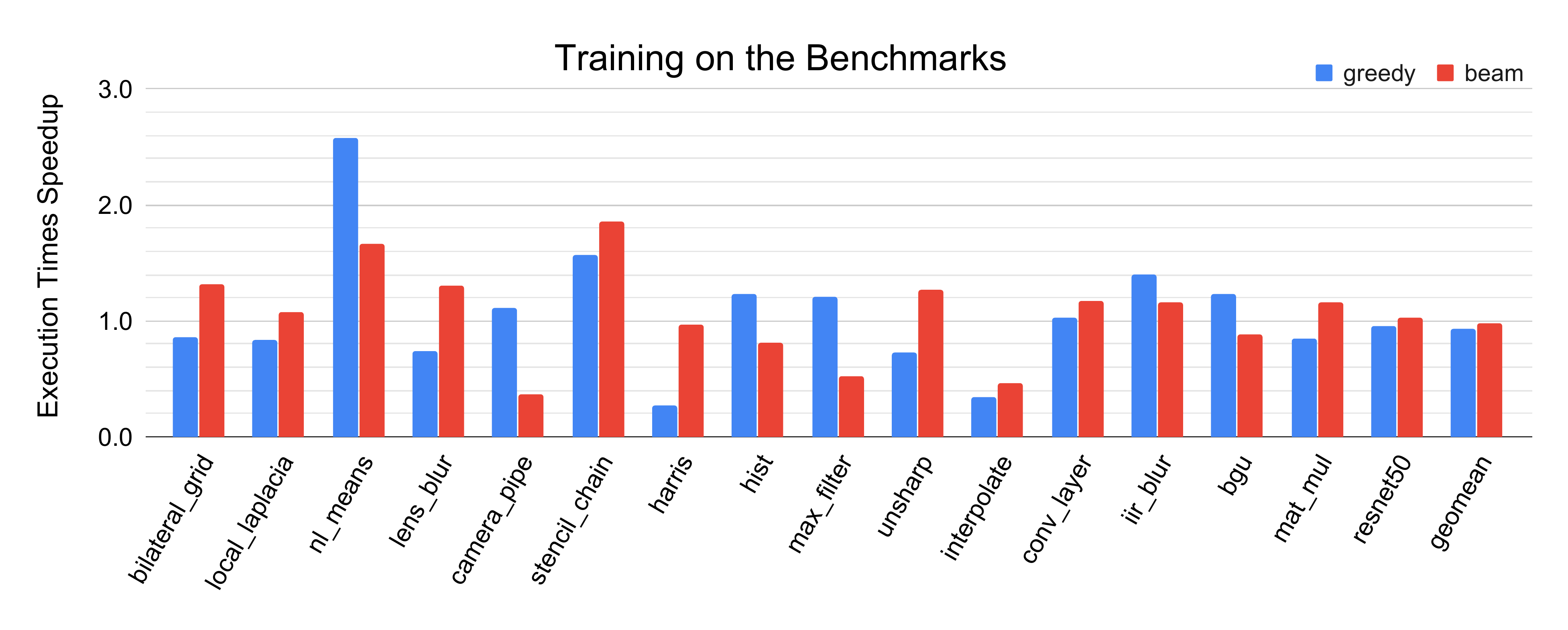}
    \caption{Speedup of greedy and beam search with a cost model trained directly on random schedules of all the benchmark algorithms themselves, normalized to greedy and beam search respectively with the original cost model (trained on complete schedules only).}
    \label{fig:trainedbenchmarks}
\end{figure*}

\section{Challenges in Beam Search}


A beam search-based approach~\cite{adams2019learning} has been recently proposed as a scheduler in Halide, which achieves state-of-the-art results. In this approach, a cost model is trained as a proxy for predicting the true execution time of intermediate schedules used by the beam search to determine which schedules to take. Unfortunately, the cost model is trained on fully scheduled programs and cannot predict the execution time of incomplete/partially-scheduled programs. We also observed that the cost model often falls short in properly predicting execution times of fully scheduled programs and thus often minimal cost does not necessarily mean optimal execution time. This makes the beam search very sensitive to inaccuracies in the cost model. Since beam search queries the cost model at every scheduling decision, this error aggregates. 

We experimented with two techniques to overcome the challenge of predicting the execution time of incomplete programs. In the first we trained a cost model as was done in~\cite{adams2019learning}, except that we trained it on the fully scheduled benchmarks that we later run the search on. Figure~\ref{fig:trainedbenchmarks} shows the results on beam search and greedy search (beam size of one) with the new cost model. We observe that the performance improves for some benchmarks while for others it deteriorates and overall the performance is similar. This was also observed by the authors in~\cite{adams2019learning} when they retrained their cost model on the specific benchmark programs that they were also autotuning. The reason is that even if the model is trained on the benchmarks that we later run the inference on, it is hard for the cost model to accurately predict the execution time of incomplete schedules during the search. In the second experiment shown in Figure~\ref{fig:valuefunc} we trained the model to predict the future cost of the current schedule. This also did not work well because there are multiple options for scheduling the rest of the program and thus the same partial schedule of a program can lead to different costs.





To overcome these challenges we formulate the scheduling problem as an MDP. In such a framework, a graph node represents an intermediate schedule/program, with edges between nodes representing potential scheduling actions. Thus, applying a particular schedule to a program could be seen as a simple graph traversal. Algorithms that solve such MDP's seek to find a node/set of actions that maximize the reward of the end state. In our use case, the reward would be the inverse of the execution time (thereby ensuring that maximizing the reward gives the fastest program).

MCTS is a promising algorithm for solving MDPs. We chose to use MCTS for four reasons: it is theoretically guaranteed to find the best node with sufficient time; its UCB formula balances the exploration of new states and exploitation of existing good states combined with using the expectation of future rewards which makes it more resilient to noise; its ability to look ahead before making a decision avoiding greediness; and the ability to make decisions based on costs of fully scheduled programs, which means the cost model can predict their execution time more accurately. Furthermore, MCTS allows us to combine real execution time measurements and the cost model's predictions to further improve performance.

\section{The Proposed ProTuner Scheduler}
Our MDP is defined by actions that correspond to intermediate scheduling decisions and states that represent intermediate schedules. The cost is the execution time of the schedule. To enumerate the possible schedules and evaluate their costs we use the same techniques used in~\cite{adams2019learning}. Given an n-dimensional tensor the scheduling is split to n stages and starting from the last stage and back to the input, a new scheduling decision is made at each stage, which proposes a new tiling and a compute and storage granularity at which to insert the new stage. The new tilings can be unrolled or spread across parallel threads or single instruction multiple data (SIMD) lanes. The costs of the complete schedules (at the end of simulation) are evaluated using a cost model trained on random programs that are fully scheduled. 
\begin{figure}[!t]
    \centering
    \includegraphics[width=0.5\textwidth,trim={12.7cm 2.5cm 7.5cm 1cm},clip]{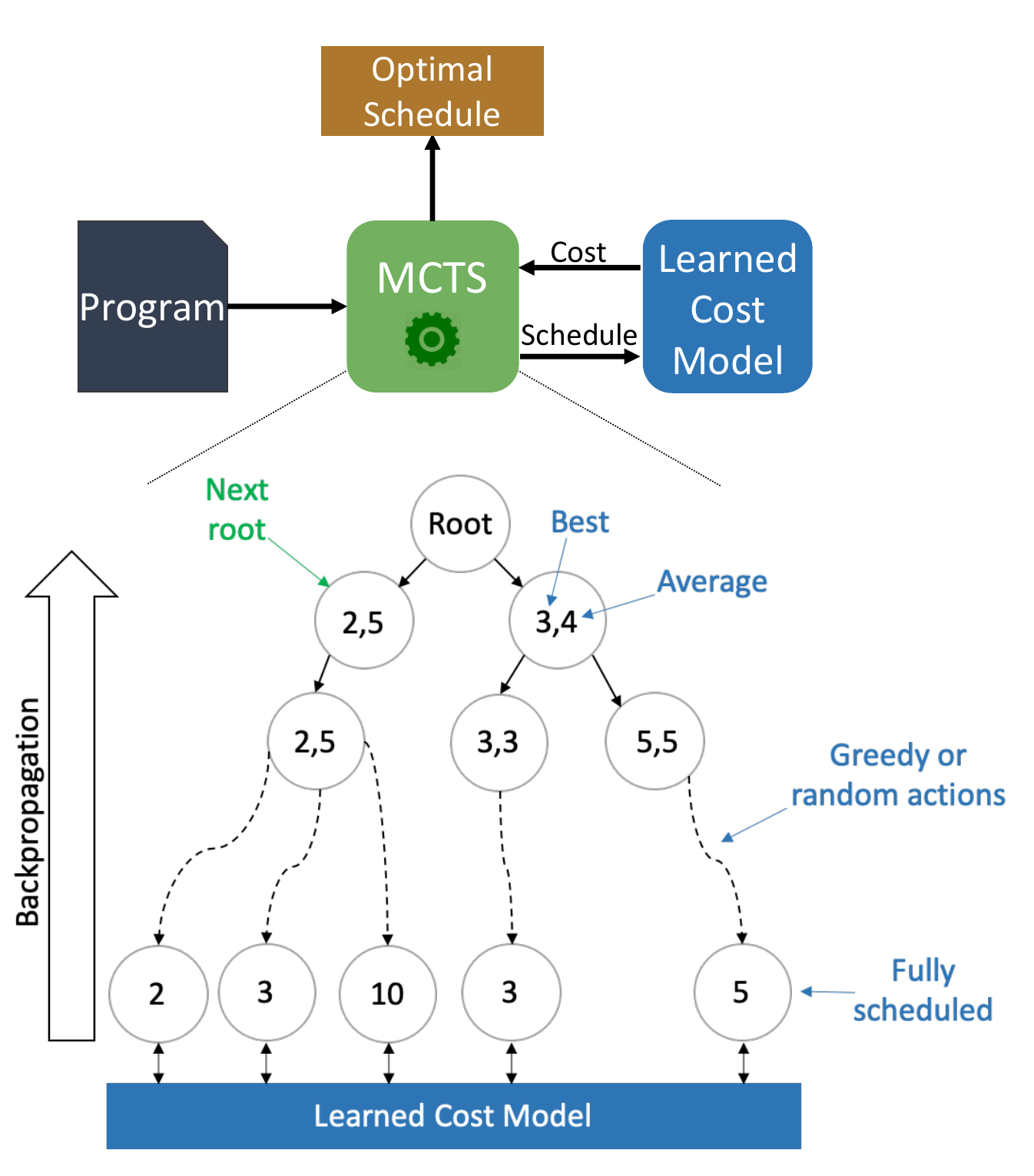}
    \caption{The block diagram of ProTuner. The program is fed to the MCTS that interacts with the learned cost model to find the optimal schedule. To make each intermediate scheduling decision the MCTS explores the benefits of the possible next actions based on the average cost but eventually picks the root that leads to the best cost. Each node stores the average costs, the best cost so far, and the complete schedule that has this best cost. The simulation can either be greedy or random. The backpropagation returns costs or 0/1 based on whether it outperforms the global best. When running an ensemble of MCTSes, the next root is picked to be the best from all the best roots.}
    \label{fig:design}
\end{figure}
\begin{figure*}[!t]
    \centering

    \begin{subfigure}[t]{0.45\textwidth}
        \centering
        \includegraphics[width=\textwidth]{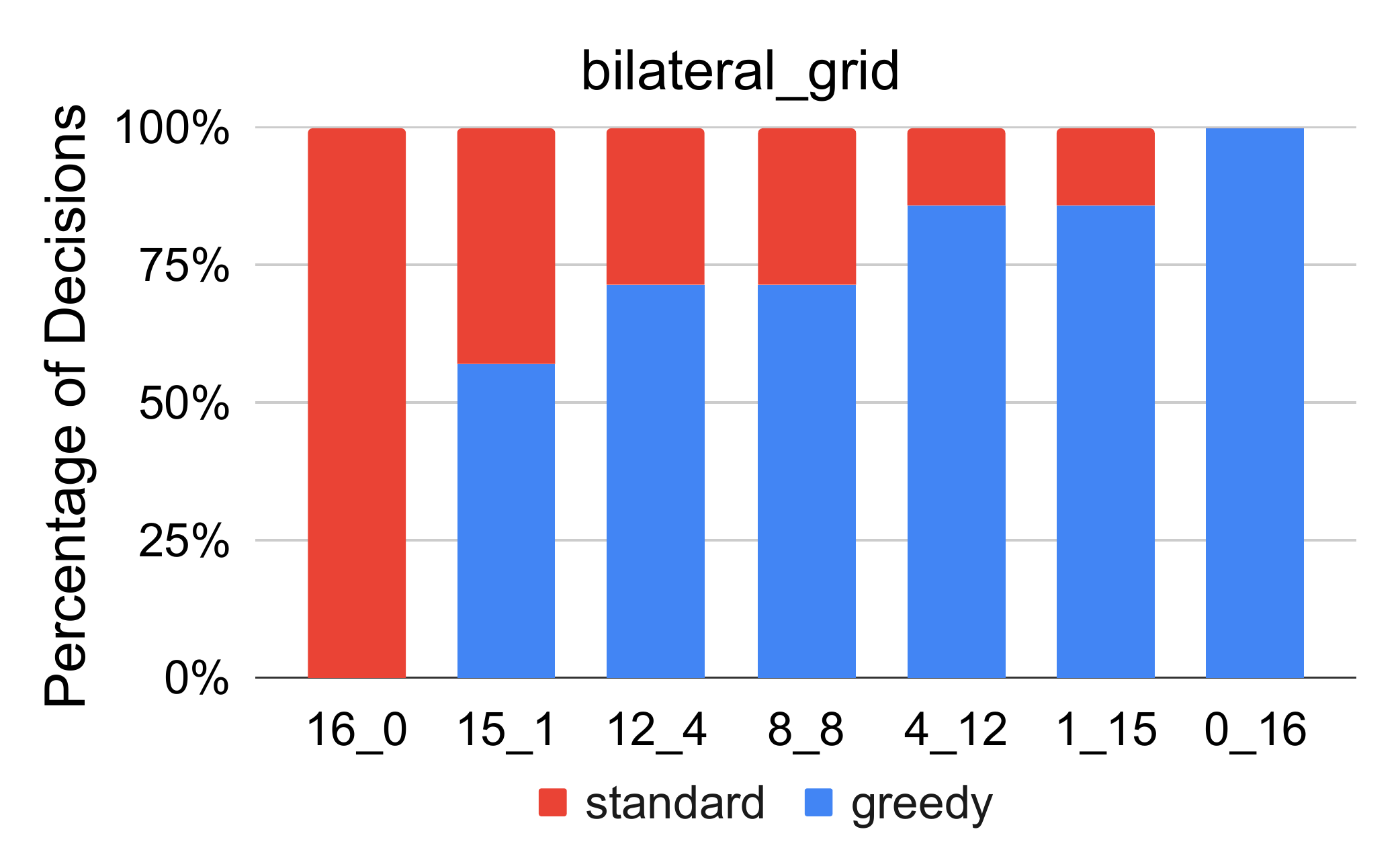}
        \caption{Proportion of decisions made by standard and greedy MCTSes on the \texttt{bilateral\_grid} test.}
    \end{subfigure}
    \hspace{0.05\textwidth}
    \begin{subfigure}[t]{0.45\textwidth}
        \centering
        \includegraphics[width=\textwidth]{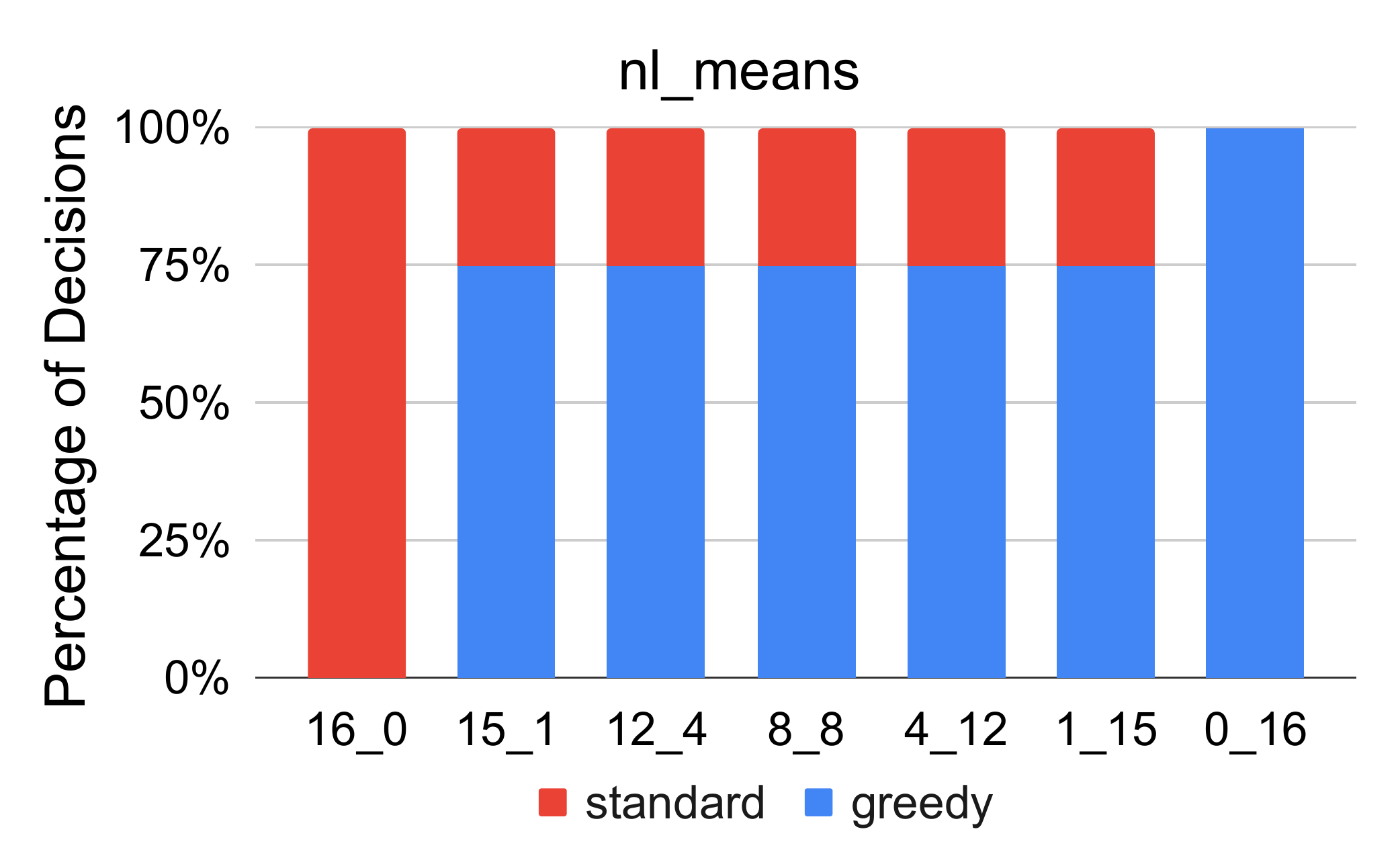}
        \caption{Proportion of decisions made by standard and greedy MCTSes on the \texttt{nl\_means} test.}
    \end{subfigure}
    
    \begin{subfigure}[t]{0.45\textwidth}
        \centering
        \includegraphics[width=\textwidth]{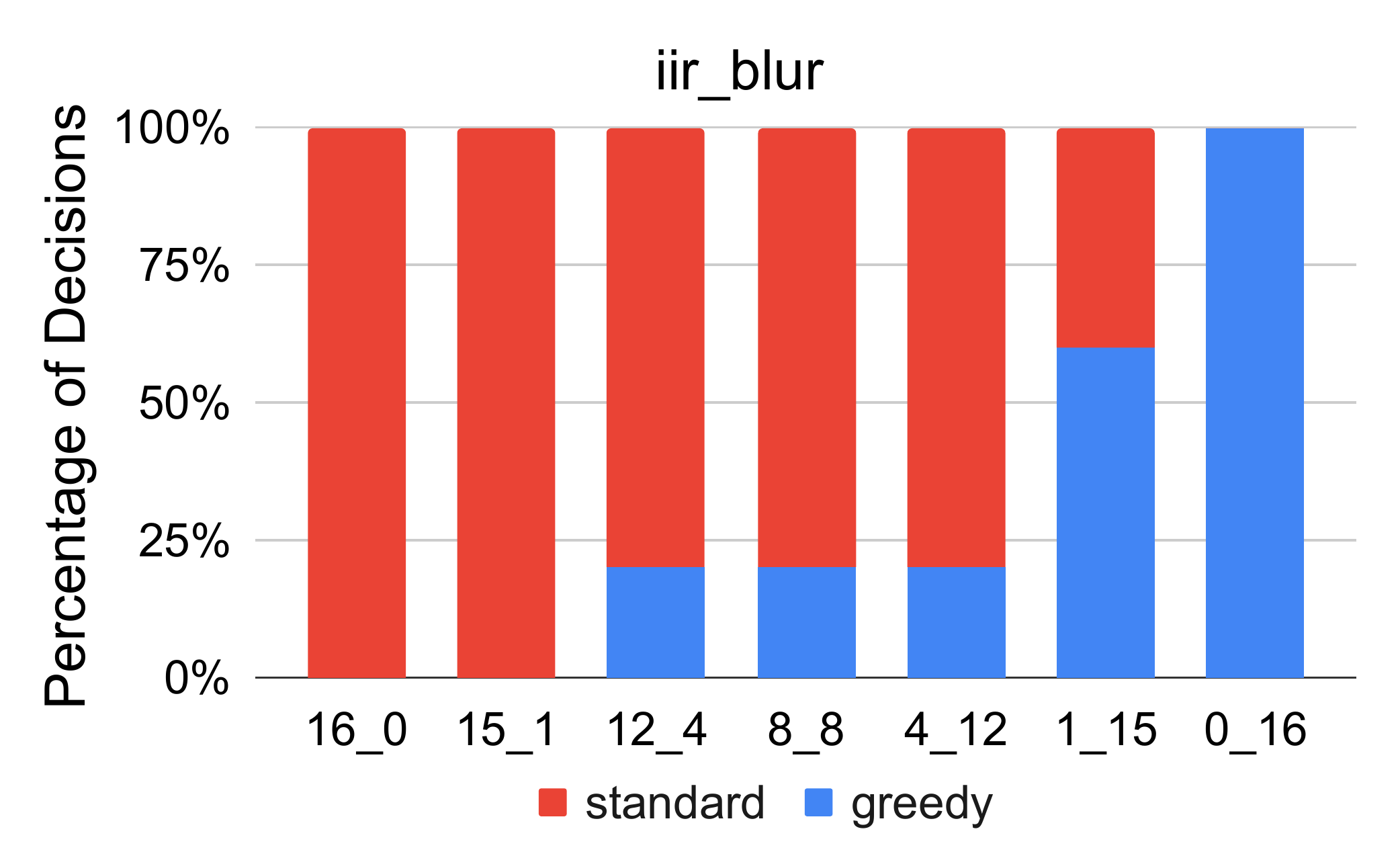}
        \caption{Proportion of decisions made by standard and greedy MCTSes on the \texttt{iir\_blur} test.}
    \end{subfigure}
    \hspace{0.05\textwidth}
    \begin{subfigure}[t]{0.45\textwidth}
        \centering
        \includegraphics[width=\textwidth]{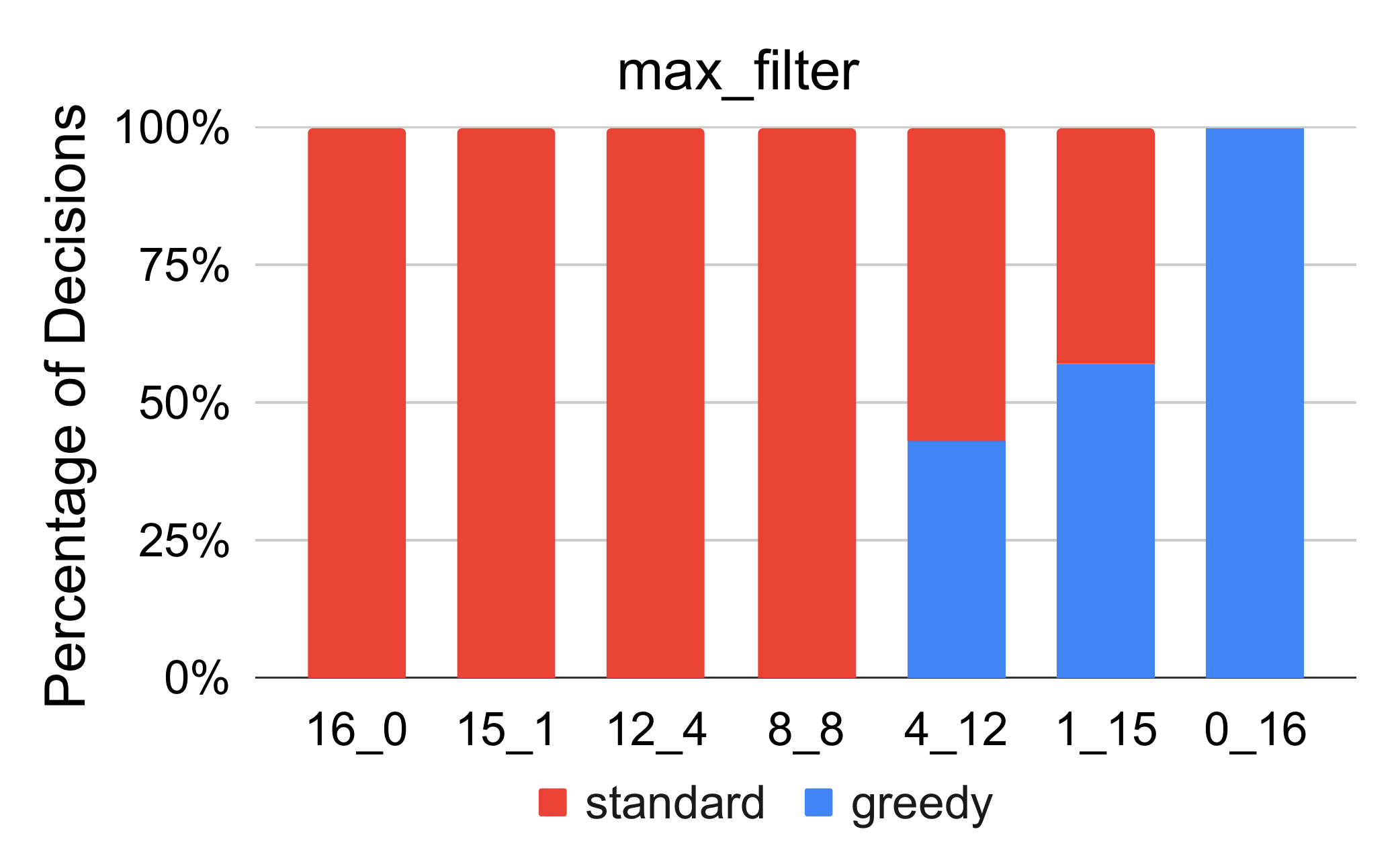}
        \caption{Proportion of decisions made by standard and greedy MCTSes on the \texttt{max\_filter} test.}
    \end{subfigure}
  
  \caption{The portion of decisions made by greedy MCTSes for a different number of standard and greedy MCTSes on a suite of four real applications. X\_Y corresponds to X standard MCTSes and Y greedy MCTSes. The overall number of trees is 16.}
  
  \label{fig:random_greedy_ratio}
\end{figure*}

Figure~\ref{fig:design} shows the block diagram of ProTuner. The MCTS starts from the last stage and explores the possible schedules back to the inputs. At every simulation from one node in the MCTS the simulation ends by computing the cost from the cost model and the cost is backpropagated to the parent nodes with the terminating fully scheduled state. These nodes update the future best cost so far, the terminating state and the value function that stores the \emph{average} cost so far. During the search the MCTS uses the average cost to determine the next child to explore. We explored the option to use the best cost in the search but that resulted in non-smooth value functions where the children that got lucky earlier and found better costs received significantly more simulations than less lucky children. This often results in a greedy behavior we are trying to avoid. 

When the computation budget is reached either after passing the number of allowed iterations or due to time out a winning action (schedule) for the current stage is determined and the new root is the state this action leads too. The winner is determined based on the \emph{best} cost so far as in \cite{bjornsson2009cadiaplayer}. We found this to outperforms taking the child with the best \emph{average} cost by 25\%. This is mainly because it can guarantee that later steps need to find schedules that are better than the best so far (rather than average best), which can be helpful when fewer iterations are available. This also allows us to combine real execution time measurements with cost model predictions at a negligible overhead.

To further improve our results we run multiple MCTSes in parallel across multiple cores that synchronize when picking a new root at every intermediate scheduling decision, which is the best child from all the best children of all the MCTSes. In addition to the performance benefits achieved from this parallelism, an ensemble of MCTSes is proven to outperform a single MCTS with the number of iterations equal to the combined number of iterations available in the ensemble~\cite{chaslot2008parallel}.

Since MCTS evaluates the program's cost when it is fully scheduled, the cost is more reliable than that of beam search, which evaluates costs of intermediate schedules that are not meaningful and the cost cannot properly evaluate since it was and could only be trained on fully scheduled programs. Unlike beam search, MCTS looks ahead and does not have the greedy nature of beam search. Furthermore, MCTS does not need to evaluate the costs of all the children during simulation or compute their state features (which we found to consume more than 92.3\% of the overhead in beam search). Instead it randomly and continuously picks a possible child and only evaluates one state when it is fully scheduled.

\subsection{Improving Scheduling Time by Adding Greedy MCTSes}
We explored multiple techniques to improve the scheduling time of our MCTS. We found that adding some greediness to our algorithm makes it find good schedules in a shorter time. First, we explored adding greediness by picking the best next action with probability $\frac{1}{2}$ instead of randomly picking an action during the simulation phase in the MCTS. This however did not give us any benefits over simulating random actions. Instead, we tried to mimic the MCTS scheme in single-player games with 0/1 rewards. So when running from the first root, it finds the best cost, and then the children that later become roots get 1 point if they beat their parent's cost, otherwise 0. This normalizes the reward, simplifies the hyperparameter tuning of the cost, and forces the new roots to beat the costs of their ancestors. However, this resulted in 9\% worse performance.

\begin{figure}[!t]
    \centering
    \includegraphics[width=0.48\textwidth]{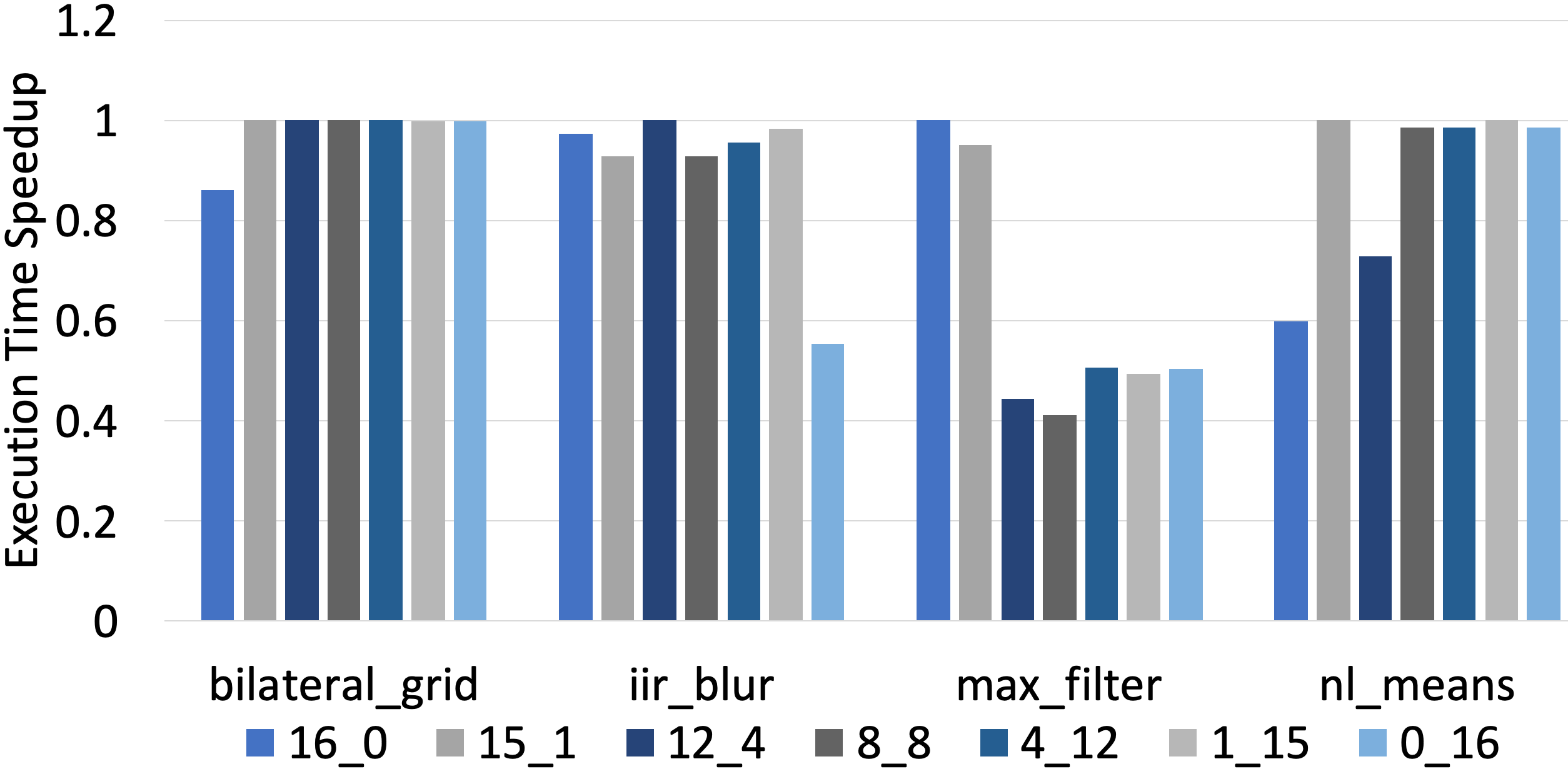}
    \caption{The execution time speed up to the best execution time on a suite of four real applications (higher is better). X\_Y corresponds to X standard MCTSes and Y greedy MCTSes. The overall number of trees is 16. The 15\_1 setting did best overall.}
    \label{fig:random_greedy}
\end{figure}

What we found to work very well was combining standard MCTSes with an MCTS that simulates greedily. In the later, after the node to be expanded based on the UCB formula is determined, it is expanded with a \texttt{random} child but the simulation is done purely greedily using the cost model. To determine how many trees should simulate randomly or greedily we experimented with different numbers of random or greedy MCTSs on four real applications as shown in Figures~\ref{fig:random_greedy_ratio} and~\ref{fig:random_greedy}. We found that some applications like \texttt{bilateral\_grid} and \texttt{nl\_means} benefited from adding greedy MCTSes while \texttt{iir\_blur} and \texttt{max\_filter} did not. We also observed that it is sufficient to use a single MCTS that simulates greedily as it made a good balance between greediness and uniform exploration. Figure~\ref{fig:random_greedy_ratio} shows the number of decisions made by greedy MCTSes as a function of different numbers of random and greedy MCTSes. We observe that adding more greedy MCTSes did not change the number of decisions made by greedy MCTSes for \texttt{nl\_means}, which benefits from greediness. For \texttt{bilateral\_grid} there was a small increase in the number of decisions by greedy MCTSes as we increased the number of greedy MCTSes but this did not impact the performance as we found that greedy MCTSes often found similar best states. For \texttt{iir\_blur} and \texttt{max\_filter} that benefit mostly from standard MCTSes, adding more greedy MCTSes made a small increase in the number of decisions made by greedy MCTSes but on the other hand the performance got worse.

\begin{figure}[!t]
    \centering
\begin{minted}{c}
all_mcts=[]
all_mcts.append(init_greedy_mcts())
all_mcts.extend(init_standard_mcts(num_mcts=15))
current_root = state0 //empty schedule
best_fully_scheduled_states={}
next_best_roots={}
while(!fully_scheduled){
  parallel_for(i=0...15){
    best_fully_scheduled_states[i], 
    next_best_roots[i] =  
        all_mcts[i].run(root=current_roots[i])
  }
  best_index = get_best_state_index_from_costs(
                    best_fully_scheduled_states)
  /* Uncomment the next lines to evaluate
  the real execution time instead of
  estimated cost:
  best_index = 
      get_best_state_index_real_measure(
        best_fully_scheduled_states) */
  parallel_for(i=0...15){
    current_roots[i] = 
        next_best_roots[best_index]
  }
  optimal_schedule = 
      best_fully_scheduled_states[best_index]
  fully_scheduled = 
      next_best_roots[best_index].is_leaf
}
\end{minted}
    \caption{Pseudocode of the MCTS scheduling algorithm that combines 15 standard MCTSes and one greedy MCTS. The best next root can be determined based on the best cost of the best fully scheduled states or based on the best execution time measurement of the best fully scheduled states as shown in the commented line.}
    \label{fig:mcts_algo}
\end{figure}

\begin{table*}[!t]
\centering
\begin{tabular}{|c|c|c|c|}
\hline
\textbf{Name} & \textbf{Seconds for Iteration} & \textbf{Expansion Formula} & \textbf{Measurement} \\ \hline
\texttt{mcts\_30s} & 30 & $\frac{1}{\frac{\sum_i ExecTime_i}{n_j}}(1 + \sqrt{\frac{ln(n)}{n_j}})$ & cost model \\ \hline
\texttt{mcts\_10s} & 10 & $\frac{1}{\frac{\sum_i ExecTime_i}{n_j}}(1 + \sqrt{\frac{ln(n)}{n_j}})$ & cost model \\ \hline
\texttt{mcts\_1s} & 1 & $\frac{1}{\frac{\sum_i ExecTime_i}{n_j}}(1 + \sqrt{\frac{ln(n)}{n_j}})$ & cost model \\ \hline
\texttt{mcts\_Cp10\_30s} & 30 & $\frac{1}{\frac{\sum_i ExecTime_i}{n_j}}(1 + 10\sqrt{\frac{ln(n)}{n_j}})$ & cost model \\ \hline
\texttt{mcts\_sqrt2\_30s} & 30 & $\frac{\sum_i\frac{1}{ExecTime_i}}{n_j} + \sqrt{2}\sqrt{\frac{2ln(n)}{n_j}}$ & cost model \\ \hline
\texttt{mcts\_cost+real\_30s} & 30 & $\frac{1}{\frac{\sum_i ExecTime_i}{n_j}}(1 + \sqrt{\frac{ln(n)}{n_j}})$ & \begin{tabular}[c]{@{}c@{}}cost model \\ + real\end{tabular} \\ \hline
\texttt{mcts\_cost+real\_1s} & 1 & $\frac{1}{\frac{\sum_i ExecTime_i}{n_j}}(1 + \sqrt{\frac{ln(n)}{n_j}})$ & \begin{tabular}[c]{@{}c@{}}cost model \\ + real\end{tabular} \\ \hline

\end{tabular}
\caption{The MCTS configurations explored. We explored different timeouts (time too determine a new root) in seconds per MCTS iteration, expansion formulas where we modify the UCB, and execution time measurement schemes. \texttt{mcts\_sqrt2\_30s} is the algorithm that gives the most weight to exploration and is the closest to the original UCB formula. \texttt{mcts\_Cp10\_30s} gives the second highest weight to exploration. \texttt{mcts\_cost+real\_30s} combines \texttt{mcts\_30s} from the first row and real execution time measurement. \texttt{mcts\_10s} and \texttt{mcts\_1s} reduce the seconds for iteration to ten seconds and one second, respectively. \texttt{mcts\_cost+real\_1s} combines \texttt{mcts\_1s} from the first row and real execution time measurement. \texttt{mcts\_sqrt2\_30s}, uses the original UCB formula, and encourages much more exploration than the other MCTS algorithms. We used $C_p = \frac{1}{\sqrt{2}}$ as suggested in~\cite{kocsis2006improved}, which showed that it works well with rewards in range $[0,1]$ as it satisfies the Hoeffding inequality. Multiplying the exploitation term with the exploration term encourages early exploitation.}
\label{tab:dse}
\end{table*}

\subsection{Combining the Cost Model and Real Execution Time Measurement}
Despite their inaccuracies, cost models are often used because the real measurement takes a prolonged time. To compensate for inaccuracies in the cost model while incurring minimal additional overhead, we added real execution time measurements at every iteration where a new root is declared. Our final algorithm is shown in Figure~\ref{fig:mcts_algo}. The algorithm initializes one greedy MCTS and 15 standard ones. While the program is not fully scheduled it runs the MCTSes in parallel from the current root. The returned best roots are evaluated based on the best real execution time measurement (the commented line).

To implement real execution time measurement in our C++ code the head thread \texttt{forks} multiple children. Each compiles a benchmark application serially for each of the candidate schedules returned by the greedy and standard MCTSes. Afterward, each of the compiled benchmark applications is run serially (to make sure they do not interfere with each other's run). The schedule with the best real execution time, rather than the one with the lowest cost, is used as the new root of the MCTSes for the next iteration.

To compile the benchmark applications, we used Halide's rudimentary ``RunGen'' wrapper, described in the official documentation~\cite{halide_rungen}. RunGen wraps a solitary scheduled Halide function (which is what we compiled) into a simple benchmark application that returns the execution time of the scheduled program. 

The RunGen wrapper fails to compile an error-free program for various applications, such as \texttt{camera\_pipe} and \texttt{bgu}. For these applications, we instead compiled the scheduled functions with the custom wrappers found under the \texttt{apps} directory in the official Halide repository. 

We could not use real execution time measurements for ResNet50 as it requires many simultaneously scheduled functions that a single forked child cannot run on its own. Therefore, for ResNet50, we report the results from the MCTS run that schedules it using the cost model only.

\begin{figure*}[!t]
    \centering
    \begin{subfigure}[t]{\textwidth}
        \centering
        \includegraphics[trim={0cm 4.5cm 0cm 0cm},clip,width=\textwidth]{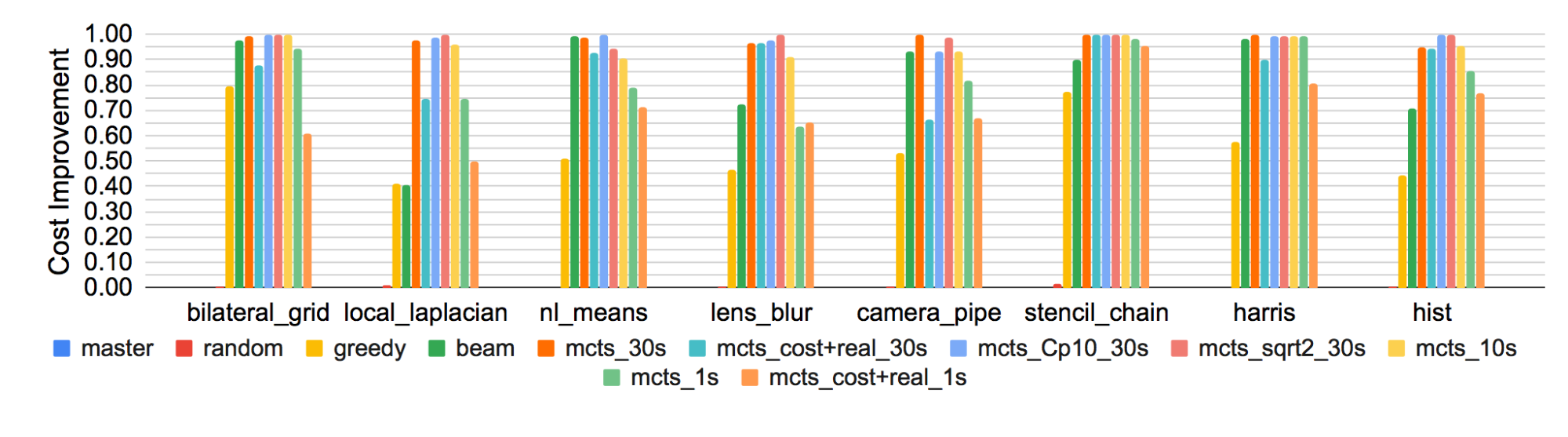}
    \end{subfigure}
    \begin{subfigure}[t]{\textwidth}
        \centering
        \includegraphics[width=\textwidth]{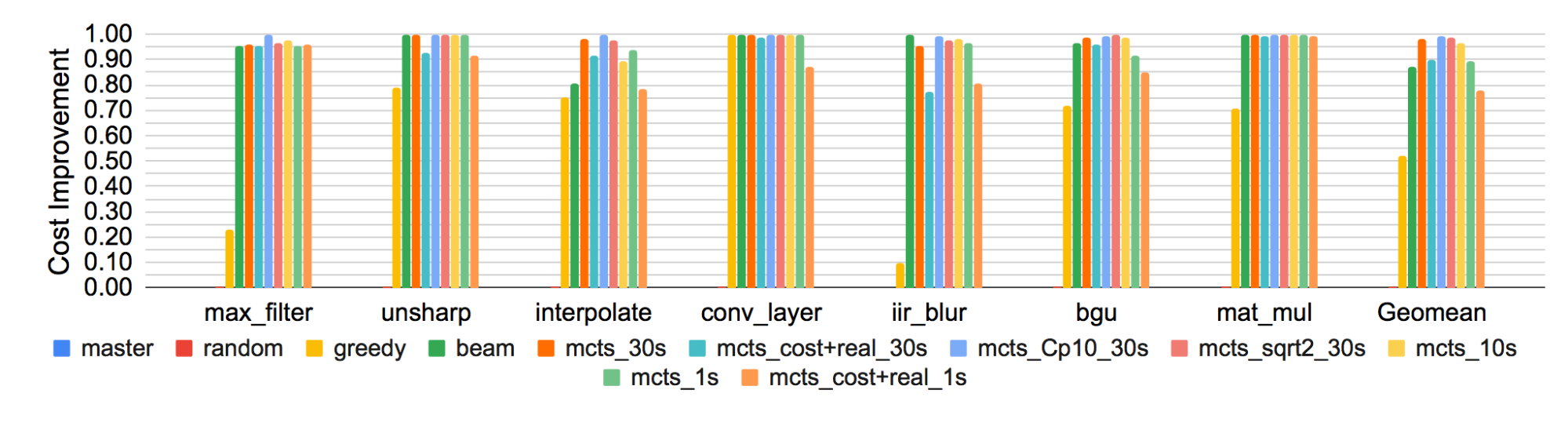}
    \end{subfigure}
    \caption{The minimum cost found by every algorithm normalized to the best cost found by all the algorithms on a suite of 16 real benchmarks.}
    \label{fig:cost}
\end{figure*}

\section{Evaluation}
\label{sec:res}


To evaluate ProTuner we built it on top of Halide in C++. We run ProTuner on AWS m5.8xlarge instances. These instances run Intel Xeon Platinum 8259CL processors with 16 physical cores with, 128 GB RAM and 100 GB SSD storage. AWS often provides different CPUs for different instances even if they are from the same type (\textit{e.g}., m5.8xlarge can have Intel Xeon Platinum 8259CL processors or Intel Xeon Platinum 8175M processors). To minimize variance between runs we run all our results on the same instance, during the same time of the day, when it is night time in the time zone and after turning off hyperthreading. Details for reproducing our results are available in the appendix. The code will also be open-sourced for further research and development.

We set the timeout limit for picking a new root in the MCTS to 30 seconds. We also explore reducing that to one second and include real execution time measurements. We use 16 MCTS (one greedy, 15 standard) that run in parallel and synchronize every timeout for picking a new root. For an apples-to-apples comparison, we also run 16 beam searches in parallel. We use the open-sourced code of Halide's beam search algorithm with the artifacts published by the original authors with the same configuration of provided in~\cite{adams2019learning} with a beam size of 32 and five passes (iterations of beam search). We also compared our results to a greedy auto-scheduler (beam size of 1), the default scheduler on Halide's master branch, and random search. Random search does not use the cost model. It runs for ten minutes and outputs the program with the best real execution time it found. The other algorithms run with three different seeds and the best performing schedule found by each algorithm is reported. 

We auto-scheduled a suit of 16 real applications. These applications range from matrix multiplications to various blurs, convolutions and interpolations, to full implementation of ResNet50~\cite{he2016deep}. These applications were taken from the baseline beam search work we compare against and are available on the Halide repository. We experimented with multiple MCTS configurations as shown in Table~\ref{tab:dse}. We mainly experimented with different timeouts for determining a new root while running the MCTS algorithm, the expansion formula that determines, which child in the tree gets expanded and integrating real execution time measurements during the search. 

\begin{figure*}[!t]
    \centering
    \begin{subfigure}[t]{\textwidth}
        \centering
        \includegraphics[trim={0cm 2.9cm 0cm 0cm},clip,width=\textwidth]{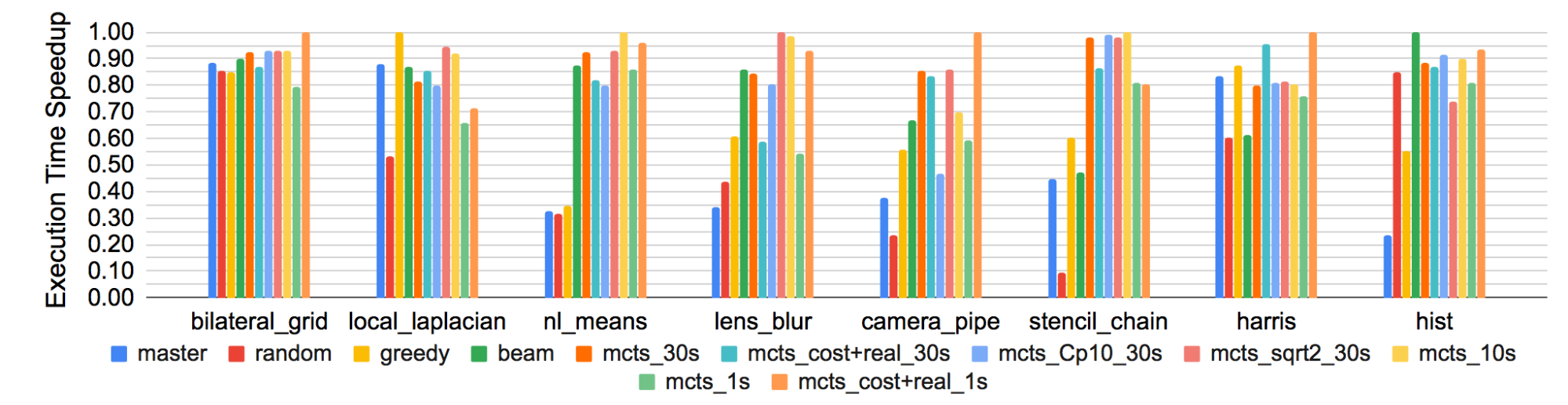}
    \end{subfigure}
    \begin{subfigure}[t]{\textwidth}
        \centering
        \includegraphics[width=\textwidth]{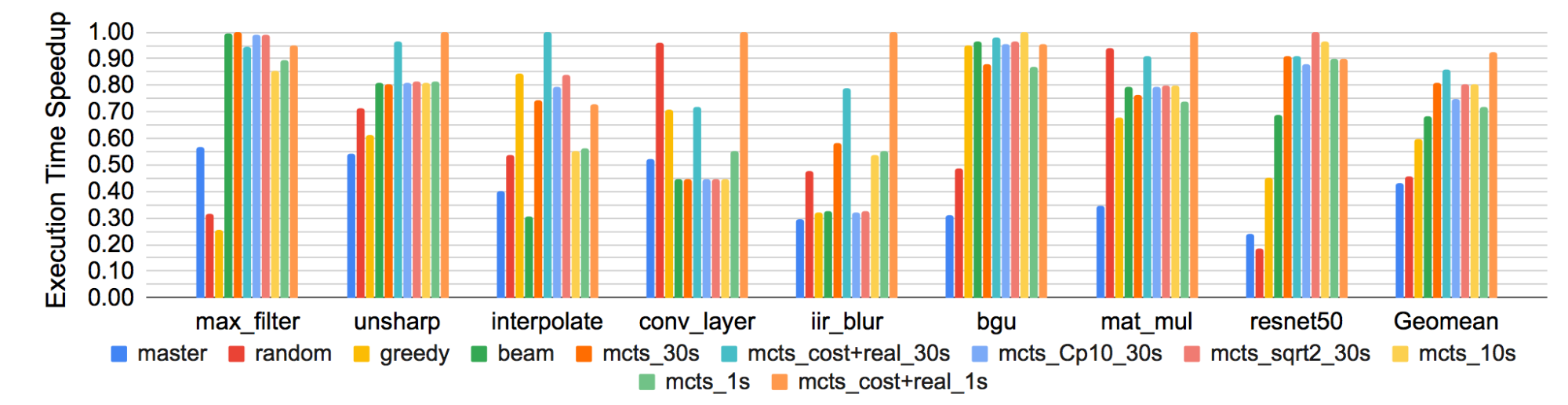}
    \end{subfigure}
    \caption{The minimum execution time found by every algorithm normalized to the best execution time found by all the algorithms on a suite of 16 real benchmarks.}
    \label{fig:perf}
\end{figure*}
\subsection{Cost}
Figure~\ref{fig:cost} shows the minimum costs found by our MCTS algorithms compared to random, greedy, and beam search. The costs are normalized to the best cost found by all the algorithms. The cost of ResNet50 is omitted because the application includes multiple stages, each stage is auto-scheduled separately (with costs in different ranges) and later the stages are merged back to form the final application.
We observe that our MCTS outperforms beam, greedy, random search cost-wise in geometric mean, in all the MCTS configurations. This means that with a cost model that has 100\% accuracy, our MCTS achieves better performance than that of beam, greedy, and random search. \texttt{mcts\_Cp10\_30s} costs outperform the costs of beam search in all the benchmarks. \texttt{mct\_30s} outperforms beam search in all the benchmarks except \texttt{iir\_blur}, in which it achieves costs 4.5\% worse than beam search. \texttt{mct\_10s} outperforms beam search in all the benchmarks except \texttt{nl\_means}, and \texttt{iir\_blur} in which it achieves costs 8.9\% and 1.1\% worse than beam search, respectively. \texttt{mcts\_sqrt2\_30s} outperforms beam search in all the benchmarks except \texttt{nl\_means} and \texttt{iir\_blur}, which achieve  costs 5.2\% and 2.4\% worse than beam search, respectively. \texttt{mcts\_cost+real\_30s} and \texttt{mcts\_cost+real\_1s} achieve the worst geometric mean cost among the MCTS algorithms. This means that they found schedules with better execution times but at higher costs. 

\subsection{Execution Time}
Figure~\ref{fig:perf} shows the minimum execution time each algorithm found normalized to the best execution time found by all the algorithms. Note that the scale in costs and execution times is different.  Similar to the costs, in geometric mean, all the MCTS algorithms outperform beam search ($1.06\times$-$1.36\times$), even \texttt{mcts\_1s}, which gives one second for each MCTS iteration. ur biggest execution time improvement is observed in \texttt{interpolate} where we achieve $1.8\times$-$3.25\times$ better performance in the different MCTS algorithms. 

As expected,  \texttt{mcts\_cost+real\_30s} and \texttt{mcts\_cost+real\_1s} achieve the best performance in geometric mean. This is despite not achieving the best geometric mean in costs. This shows that real execution time measurement is effective. Interestingly, \texttt{mcts\_cost+real\_1s} achieves better performance than \texttt{mcts\_cost+real\_30s}. The first is less likely to overfit to the cost model than the second. A clear example can be seen in \texttt{conv\_layer}, which is the smallest benchmark. While both \texttt{mcts\_cost+real\_1s} and \texttt{mcts\_cost+real\_30s} outperform the other MCTS algorithms, the first achieves better performance as it stops evaluating the cost model earlier and hence it is less likely to overfit to the cost model, especially on smaller benchmarks.


We observe that \texttt{mcts\_10s} achieves similar performance to \texttt{mcts\_30s} and \texttt{mcts\_sqrt2\_30s}, and better performance than \texttt{mcts\_Cp10\_30s} in geometric mean. This means that 10 seconds per MCTS iteration is sufficient for our benchmarks. Adding more time might be useful for large benchmarks but can be harmful to smaller benchmarks where the MCTS might overfit to the cost model. An ideal configuration should consider the size of the application when setting the MCTS parameters.

\begin{figure*}
    \centering
    \includegraphics[width=\textwidth]{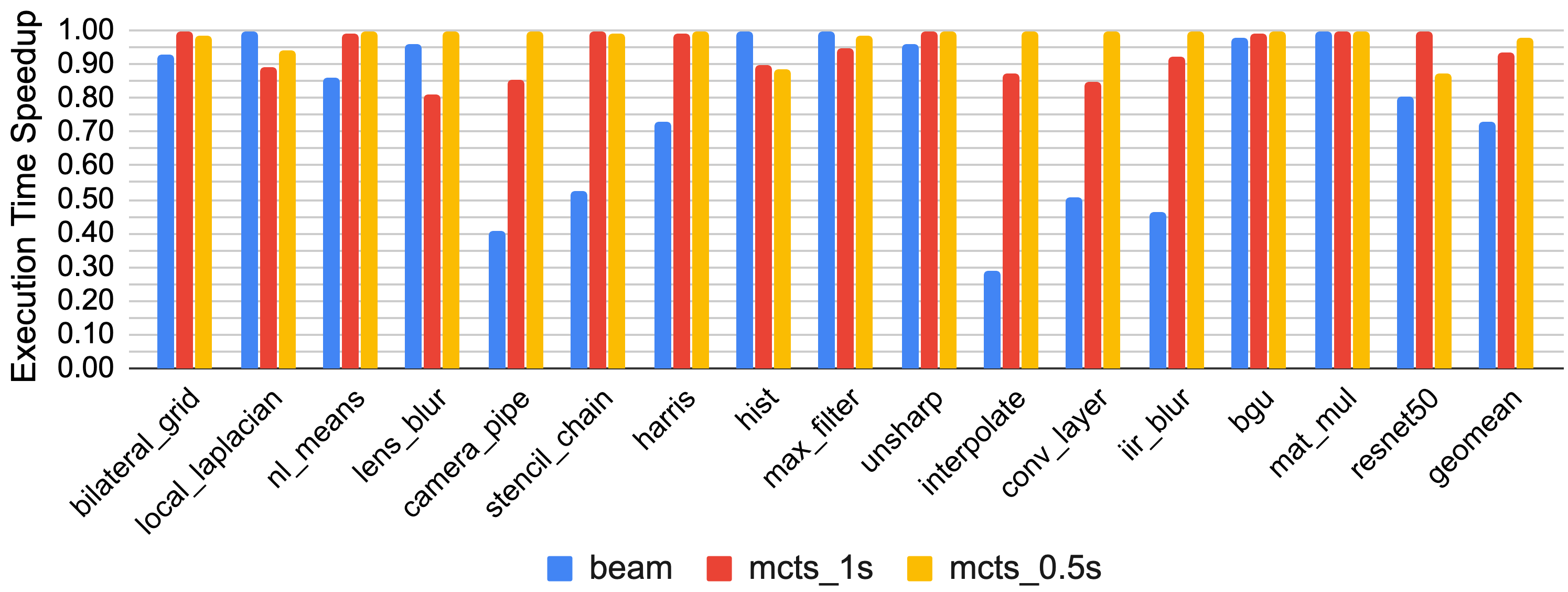}
    \caption{Execution time speedup normalized to the best execution time using beam search, \texttt{mcts\_1s}, and \texttt{mcts\_0.5s} with autotuning on a suite of 16 real benchmarks. Each algorithm is rerun with a different seed until a timeout of 15 minutes is reached and the bests performance found by each algorithm is reported.}
    \label{fig:autotune}
\end{figure*}


\subsection{Search Time}
Our MCTS algorithms with the cost model schedule programs in seconds to minutes. Among all the benchmarks, the average auto-scheduling time of \texttt{mcts\_1s}, \texttt{mcts\_10s}, and \texttt{mcts\_30s} are 31, 155, 422 seconds respectively. This includes the time to compile the search code, the search time, and the time to compile and benchmark the applications. In smaller applications, such as~\texttt{conv\_layer} and~\texttt{mat\_mul}, this time is dominated by the compilation and benchmarking time. For larger applications, this time is dominated by the search time. Our performance analysis shows that most of the search time (88\%) is spent during the generation of new children (schedules) in the simulation phase and only 7.5\% of the time is spent in the cost evaluation. However, our standard MCTS simulation needs a single randomly generated child. The rest of the children are generated but not used. Therefore we see a potential for $8\times$ speedup in the search time of MCTS, which we seek to implement in future work.

For \texttt{mcts\_cost+real\_1s} and \texttt{mcts\_cost+real\_30s} the average auto-scheduling time is 23 and 35 minutes, respectively. Most benchmarks require roughly $3\times$ more time to auto-schedule with real execution time measurement when the MCTS iteration is set to 30 seconds. This time mostly consumed in the forked children processes that compile and evaluate the candidates of potential next roots serially. While it might seem to be inefficient to do this serially, we chose do it that way to avoid interference between threads during execution time measurement. 


\subsection{Autotuning with Limited Time Budget}
Figure~\ref{fig:autotune} shows the autotuning performance comparison between beam search and MCTS. We limit the autotuning time to 15 minutes for each application. This time includes the compilation and execution time of the benchmarks as well as the search time. The execution time of the generated program is normalized to the best execution time found by beam search and MCTS. For time efficiency we use \texttt{mcts\_1s}. We further explored using half a second per MCTS iteration (\texttt{mcts\_0.5s}), which allows for more real execution time measurements during autotuning. Each algorithm is rerun with a different seed until the timeout of 15 minutes is reached and the best performance found by each algorithm is reported. \texttt{mcts\_0.5s} achieves the best overall performance and outperforms beam search by up to $3.43\times$ and by $1.35\times$ in geometric mean, in the same time budget. \texttt{mcts\_1s} also outperforms beam search by up to $3\times$ and by $1.29\times$ in geometric mean in the same time budget. This is mainly due to more accurate scheduling decisions derived by meaningful costs of fully scheduled programs rather than incomplete programs. 

\section{Related Work}
Multiple previous attempts to automatically schedule Halide programs have been proposed. In the original paper~\cite{ragan2013halide}, heuristics and genetic search over random schedule rewrites were used. The scheduling relied on measuring real execution time, which resulted in a search time that can get prolonged to days for moderate benchmarks. OpenTuner~\cite{ansel2014opentuner} autotunes a program using an AUC-Bandit-meta-technique-directed ensemble selection of algorithms. It is effective in scheduling simple pipelines \cite{mullapudi2015polymage}.

The auto-scheduler that comes with the master Halide repo, which we compare against in Section~\ref{sec:res} is based on~\cite{mullapudi2016automatically}. It uses a cost model with a greedy search algorithm that allows it to run quickly, with not autotuning or benchmarking at all. However, it only considers a fixed set of optimization heuristics for things like parallelism, vectorization and unrolling, and a single level of tiling and fusion. \cite{jangda2018effective} improved this cost model, but did not expand the restricted search space. \cite{sioutas2018loop,sioutas2019schedule} improves the search space and uses a manual cost model. However, the search space is still smaller than ours. 

Many compilers that use loop polyhedral analysis to perform automatic scheduling of affine loop nests~\cite{mullapudi2015polymage,vasilache2018tensor,baghdadi2015pencil,bondhugula2008practical,grosser2012polly,baghdadi2019tiramisu}. Many possible Halide schedules are excluded in these compilers. However, it might be possible to use the polyhedral representation in building more accurate cost models, which we plan to do in future work.

AutoTVM~\cite{chen2018learning} uses tree-based algorithms to auto-schedule programs on TVM~\cite{chen2018tvm}, which is an optimization stack for deep learning. This approach however still requires the user to manually write the search space for each loop. Furthermore, each operation is optimized in isolation without exploring large programs. \cite{ahn2019reinforcement} uses deep reinforcement learning to schedule deep learning pipelines and improves the performance compared to AutoTVM.

Machine learning in compiler optimization has been proposed in many prior works~\cite{wang2018machine,ashouri2018survey}. This includes phase ordering~\cite{fursin2008milepost,huang2013effect,huang2015effect,agakov2006using,2003Stephenson,2012Kulkarni}, tiling factors~\cite{rahman2010neural}, mappings of kernels to CPUs or GPUs~\cite{cummins2017end} with supervised learning, auto-vectorization~\cite{haj2020neurovectorizer,stock2012using,tian2016llvm,trifunovic2009polyhedral,nuzman2011vapor,porpodas2015throttling,larsen2000exploiting,mcfarlin2011automatic,porpodas2017supergraph,porpodas2015pslp,mendis2019compiler} and the throughput of basic blocks~\cite{mendis2018ithemal}.

Multi-Level Intermediate Representation (MLIR)~\cite{lattner2020mlir} has been recently proposed to help with scaling the performance with the end of Moore's law. One objective of MLIR is to represent kernels in a form suitable for optimization, and allow easy integration of search algorithms such as reinforcement learning, MCTS, and beam search.

\section{Future Work}
We see multiple future directions for this work. We could combine a value function cost model that can predict the advantage of taking an action instead of running the MCTS simulation. Another direction can include applying deep reinforcement learning methods to solve the scheduling MDP as done in similar domains in compiler optimization in~\cite{haj2020autophase,haj2020neurovectorizer,huang2019autophase, ahn2019reinforcement}. We believe that if deep reinforcement learning (with a neural network) can generalize in that case then the runtime of the algorithm can be significantly improved as the algorithm will only need to run inference rather than retrain/research from scratch as the case in MCTS or beam search.
With limited resources, more accurate cost models are necessary for scheduling,  especially with the recent trends in customized hardware and the explosion of new applications. Such cost models need to consider the target hardware parameters and program features.

Another point for improvement in our implementation is in generating the next states during the random simulation. During the MCTS simulation phase, our implementation now generates all the possible children (next possible intermediate schedules) and then randomly picks one child. In this setting, our algorithm's cost evaluation overhead is 7.5\%  while 88\% of the time is spent on enumerating children that our standard MCTSes do not use. In other words, we see the potential for 8$\times$ runtime improvement for our MCTS algorithm. Different configurations of the MCTS impact the performance differently. Some applications seem to benefit more from greedy behavior while others work better with a random one. The $C_p$ could be further tuned and our performance could be improved if we had a different $C_p$ for different programs. This is because different programs have different costs/runtimes and thus using the same $C_p$ for all programs encourages less exploration in shorter programs. Furthermore, as we go deeper into the MCTS, the $C_p$ could be reduced to encourage less exploration as the standard deviation in the costs of the children decreases. We also observed that our MCTS can overfit to the cost model if we run it for too long. 

We plan to extend the search space to include more optimizations currently pruned such as vector sizes different than the native vector size, or not enforcing the multi-core parallelism to be at the outermost loop level. We also plan to explore more hardware targets and experiment with more autotuning methods that can further improve the performance.
\section{Conclusion}
We proposed and developed ProTuner: a framework that uses an MCTS-based algorithm to automatically tune programs for high-performance deep learning and image processing applications. Our results demonstrated up to $3.25\times$ better performance compared to the state-of-the-art beam-search algorithm. Looking forward, we foresee a potential opportunity to continue scaling performance---despite the end of Moore's Law---through automatic program tuning and optimization, with machine learning algorithms such as MCTS.

\bibliographystyle{plain}
\bibliography{main.bib}

\clearpage

\section*{Appendix}
\subsection{Running}
Once the Github repository is cloned and the machine is setup run:
\begin{minted}{bash}
./generate_all_apps_results.sh
\end{minted}
inside Halide/apps/autoscheduler/paper\_results\_scripts directory.
\subsection{Setting AWS Instance}
we use Ubuntu Server 18.04 LTS (HVM), SSD Volume Type - ami-003634241a8fcdec0 (64-bit x86)  m5.8xlarge instances with 100GB Storage.
\subsubsection{Turning Off Hyperthreading}

\begin{minted}{bash}
#!/bin/bash
for i in {16..31}; do
  echo "Disabling logical HT core $i."
  echo 0 > /sys/devices/system/cpu/cpu${i}/online;
done
\end{minted}
\subsubsection{Set Up}
\begin{minted}{bash}
sudo apt-get install libcurl4-openssl-dev libssl-dev
uuid-dev zlib1g-dev libpulse-dev -y
#Might need :
#sudo apt-get clean
#sudo apt-get update
sudo snap install cmake --classic
git clone https://github.com/llvm/llvm-project.git
cd llvm-project
git checkout release/9.x  # to build LLVM 9.x
mkdir build
mkdir install
cd build
cmake -DLLVM_ENABLE_PROJECTS="clang;lld;clang-tools-extra" 
-DLLVM_ENABLE_RTTI=ON -DLLVM_ENABLE_TERMINFO=OFF
-DLLVM_TARGETS_TO_BUILD="X86;ARM;NVPTX;AArch64;Mips;PowerPC;Hexagon" 
-DLLVM_ENABLE_ASSERTIONS=ON -DCMAKE_BUILD_TYPE=Release
-DLLVM_BUILD_32_BITS=OFF -DCMAKE_INSTALL_PREFIX=../install 
../llvm
make install -j16
export LLVM_CONFIG=~/llvm-project/install/bin/llvm-config
cd ..
git clone https://github.com/Anonymous/Halide
cd Halide
make -j30
install anaconda
curl -O https://repo.anaconda.com/archive/Anaconda3-2019.03-Linux-x86_64.sh
sha256sum Anaconda3-2019.03-Linux-x86_64.sh
bash Anaconda3-2019.03-Linux-x86_64.sh
source ~/.bashrc
conda create --name halide python=3
conda activate halide
pip install torch torch-vision numpy
conda install -c anaconda libpng
sudo apt-get install libjpeg-dev
export LD_LIBRARY_PATH=~/llvm-project/Halide/bin
cd ~/llvm-project/Halide/apps/autoscheduler/\
paper_results_scripts
./generate_all_apps_results
\end{minted}

\end{document}